\documentclass[prc,showpacs,preprintnumbers,unsortedaddress,amsmath,amssymb,floatfix]{revtex4}
\usepackage{graphicx,color}
\usepackage{bm}
\usepackage{graphicx}
\usepackage{epsf}

\usepackage{comment}
\usepackage{color}
\usepackage{slashed}
\usepackage{ulem}
\begin{document}

\title{Medium polarization in asymmetric nuclear matter}

\author{S. S. Zhang$^{a}$, L. G. Cao$^{b,c}$, U. Lombardo$^{d}$
\footnote{Corresponding author at: Laboratori Nazionali del
Sud (INFN), via S. Sofia 62, 95123 Catania, ITALY, phone:
+39.095.542.111, fax: +39.095.71.41.815, email: lombardo@lns.infn.it.
}, P. Schuck$^{e,f}$}

\affiliation{$^{a}$ School of Physics and
Nuclear Energy Engineering, Beihang University, Beijing 100191,
China}

\affiliation{$^{b}$ School of Mathematics and Physics, North China
Electric Power University, Beijing 102206, China}

\affiliation{ $^{c}$ State
Key Laboratory of Theoretical Physics, Institute of Theoretical
Physics (CAS), Beijing 100190, China}

\affiliation{$^{d}$ Dipatimento di
Fisica e Astronomia dell'Universit\'a and Laboratori Nazionali del
Sud(INFN), via S. Sofia 62, 95123 Catania,Italy}

\affiliation{ $^{e}$
Institut de Physique Nucl\'eaire, Universit\'e Paris-Sud, F-91406
Orsay Cedex,  France  }

\affiliation{$^{f}$ Laboratoire de Physique et de Mod\'lisation
des Milieux Condens\'es, CNRS and Universit\'e Joseph Fourier, 25 Av. des Martyrs, BP 166, F-38042 }

\date{\today}

\begin{abstract}
The influence of the core polarization on the effective nuclear
interaction of asymmetric nuclear matter is calculated in the
framework of the induced interaction theory. The strong isospin
dependence of the density and spin density fluctuations is studied
along with the interplay between the neutron and proton core
polarizations. Moving from symmetric nuclear matter to pure neutron
matter the crossover of the induced interaction  from attractive to
repulsive
 in the spin singlet state is determined as a function of the isospin imbalance.
 The density range in which it occurs is also determined.
For the spin triplet state the induced interaction turns out to be
always repulsive.
  The implications of the
results for the neutron star superfluid phases are shortly
discussed.
\end{abstract}

\pacs{21.30.Fe,21.65.Cd,26.60.-c}
\maketitle

\section{Introduction}
Recently the  interest for the superfluid states of neutron stars
(NS) has been reviving after the real time temperature measurements
of CasA remnant \cite{CasA}. Theoretical models, devised to explain
the cooling of such a system, demand for accurate predictions of the
pairing gaps\cite{cooling}. In particular, the study was focussed on
the $^1S_0$ proton-proton (p-p) and the $^3PF_2$ neutron-neutron
(n-n) pairing in the NS core. The peculiar aspect of NS pairing is
that the onset of superfluidity occurs at high nuclear density,
where the $\beta$-stability condition imposes large isospin
imbalance. The latter feature could in fact influence to a large
extent the vertex corrections to be included in the gap equation.
Whereas it was definitely established that the self-energy
corrections suppress the pairing magnitude\cite{dong}, to what
extent the core polarization affects the pairing interaction is not
yet sufficiently clarified. This effect has been studied for several
years (an extensive bibliography is in Ref.\cite{ulom}) in the
framework of the induced interaction theory (IIT)\cite{babu,sjob}.
More recently, Ref.\cite{cao}considered the two extreme situations
of pure neutron matter(PNM) and symmetric nuclear matter(SNM). It
was found that the core polarization, that in PNM quenches the n-n
$^1S_0$ pairing gap, on the contrary enhances it in SNM. The main
conclusion was that in SNM a large compensation occurs between
self-energy and vertex corrections, at least in the $^1S_0$ n-n
pairing. More recently polarization effects based on the RPA limit
were included into the interaction to calculate the $^1S_0$ p-p
pairing in $\beta$-stable nuclear matter\cite{baldo} and the
$^3PF_2$ n-n pairing in pure neutron matter\cite{dickhoff}.

The present paper is intended to extend the study 
of Ref.\cite{cao}
to asymmetric nuclear matter and to report new calculations of the
particle-hole (ph) residual interaction and the polarization
propagator, in the realistic context of $\beta$-stable nuclear
matter suitable for application to the NS pairing. The main scope is
to investigate the interplay between density fluctuations and spin
density fluctuations in singlet and triplet spin states within the
particle-particle (pp) coupling, and in particular the crossover
from attractive to repulsive interaction due to the core polarization,
when moving from SNM to PNM. The baryon density threshold $\rho_c$,
at which such a mechanism disappears, is also determined. These
pieces of information are essential to study the various NS
superfluid states.

    \section{Formalism}

    The induced interaction theory for symmetric nuclear matter is reviewed in
    Ref.\cite{back} (see also Ref.\cite{cao} and references therein quoted).
    Its extension to asymmetric nuclear matter is straightforward. In this case the ph
    irreducible interaction $\cal{J}^S_{\tau\tau'}$ is still  rotationally invariant
    in spin space, S being  the total spin in ph coupling, but it is not in
    isospin space, i.e. ${\cal{J}}^S_{nn} \neq {\cal{J}}^S_{pp}$.
    In the IIT framework $\cal{J}^S_{\tau\tau'}$
    is determined by the equation diagrammatically depicted  
    
\begin{figure}[htb]
\centering
\vspace*{-2.in}
\includegraphics[angle=0,width=12cm]{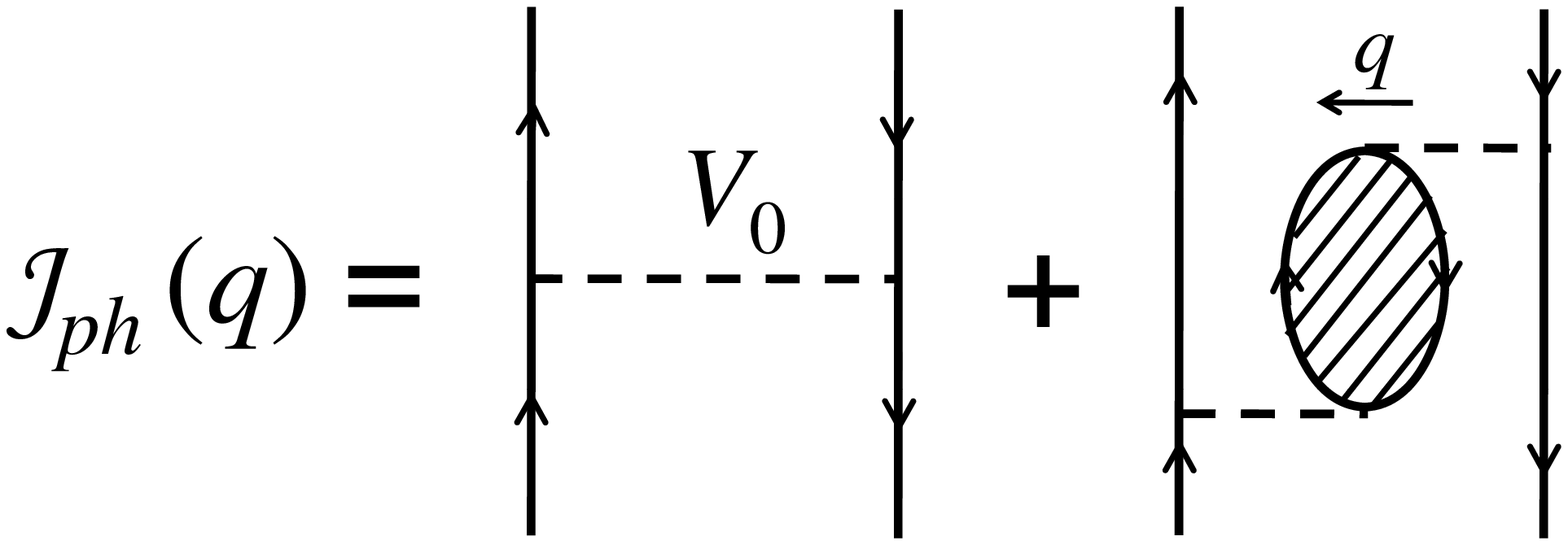}
\caption{Particle-hole induced interaction. The dotted line
represents the direct term (G-matrix in the present
approximation), the dashed lines represent the ${\cal{J}}_{ph}$
itself, and q is the momentum transfer.}\label{diagram}
\end{figure}

\noindent in Fig.1:
    The driving term (first term on the r.h.s. of Fig.1) is assumed to be approximated by G-matrix
    ; the inducedterm (second term on the r.h.s.) describes the
    interaction mediated by the medium excitations via
    the dressed polarization propagator (bubble in Fig.1). The vertex insertions represent
     the full ph irreducible
    interaction itself. The analytic expression of $\cal{J}^S_{\tau\tau'}$ is given
    by the 2$\times$2 matrix equation in isospin space

    \begin{eqnarray}
    {\cal{J}}^S(q)  &=&  G^S(q) + {\cal{J}}_i^S(q) \\
    {\cal{J}}_i^0(q)&=& \frac{1}{2}\sum_{S'}
    (2S'+1){\cal{J}}^{S'}(q)\Lambda^{S'}(q){\cal{J}}^{S'}(q) \\
    {\cal{J}}_i^1(q)&=& \frac{1}{2}\sum_{S'}
    (-1)^{S'}{\cal{J}}^{S'}(q)\Lambda^{S'}(q){\cal{J}}^{S'}(q),
    \end{eqnarray}
    where $q\equiv(\vec q,\omega)$ is the energy-momentum  transfer and
    ${\cal{J}}_i$ denotes the induced
    interaction. This equation is solved in the Landau limit \cite{sjob}. The polarization propagator
    $\Lambda^S(q)$ is calculated as bubble series but, at variance with
    RPA, the vertex insertions between bubbles are expressed in terms of whole ph
    interaction  ${\cal{J}}^S$\cite{babu}. In that limit this
    series can be summed up, resulting into  a simple algebraic
    formula for the diagonal matrix elements
    \begin{eqnarray}
    \Lambda^S_{\tau,\tau}(q)&=&\lambda_{\tau}(q)\frac{(1+\lambda_{\tau'}(q){\cal{J}}^S_{\tau',\tau'}(q))}{{\cal{D}}^S(q)}
    \end{eqnarray}
    and for the off-diagonal matrix elements
    \begin{eqnarray}
    \Lambda^S_{\tau,\tau'}(q)&=&-\frac{\lambda_{\tau}(q)\lambda_{\tau'}(q){\cal{J}}^S_{\tau,\tau'}(q)}{{\cal{D}}^S(q)}
    \\
    1/{\cal{D}}^S &=& (1+\lambda_{\tau}{\cal{J}}^S_{\tau,\tau})(1+\lambda_{\tau'}{\cal{J}}^S_{\tau',\tau'})-\lambda_{\tau}
    \lambda_{\tau'}({\cal{J}}^S_{\tau,\tau'})^2 \,,
    \end{eqnarray}
    where $\tau\ne\tau'$. $\lambda_{\tau}(q)$ is the free polarization propagator\cite{fetter} corrected by the
    mean-field effects (in the last line the q dependence is understood). Its explicit form is
    \begin{equation}\nonumber
    \lambda_{\tau}(q)\,=\,Z^2_{\tau}N_{\tau}\lambda(\frac{k}{k^F_{\tau}},\frac{\omega}{\varepsilon^F_{\tau}}),
    \end{equation}
    where $\lambda(q)$ is the Lindhard function\cite{lind}. $N_{\tau}$ and $Z_{\tau}$ are level density
    and quasi-particle strength at the Fermi
    surface, respectively, calculated within the extended  Brueckner-Hartree-Fock (BHF) approximation (see
    Ref.\cite{dong}). It is worthwhile noticing  that the bubble expansion is expected to
    be rapidly convergent at high density because $Z^2<<1$.

\section{Results and discussion}

As in Ref. \cite{cao}, Eqs.(1)-(3) have been solved by means of the
so called LNS potential, that is a Skyrme parametrization of the
energy density functional calculated in the BHF 
approximation\cite{lns}. The difference from the pure Skyrme force
is that the LNS parameters are derived from $\it ab\, initio$ calculations
rather than by fitting the empirical nuclear data. The LNS interaction
enables a strong simplification of the formalism
\cite{giai,navarro}.

\begin{figure}[htb]
\centering
\includegraphics[angle=0,width=15cm]{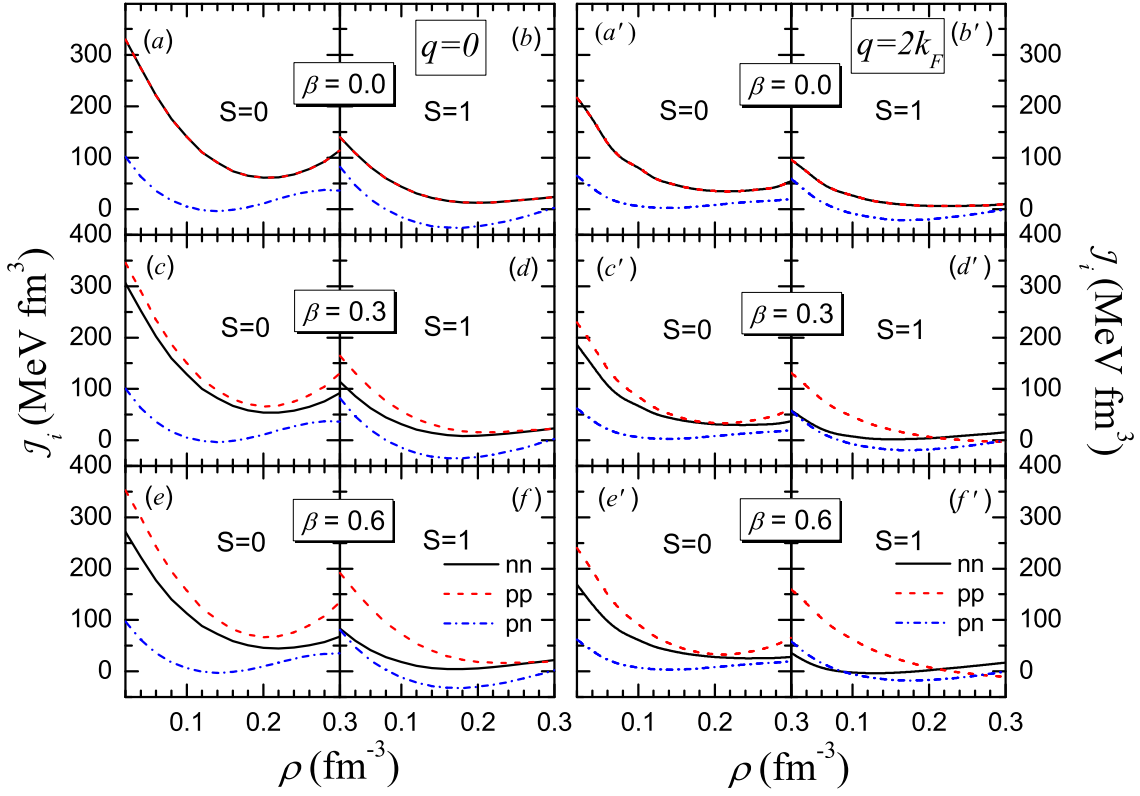}
\caption{Particle-hole induced interaction in nuclear matter vs.
density for three values of $\beta$ at $q = 0$ (left) and $q=2k_F$
(right). }\label{vph}
\end{figure}

The calculation of the ph induced interaction in nuclear matter
 $({\cal{J}}_i)_{\tau,\tau'}^S$ was performed in a wide density range at $q=0$,
 and extended to non vanishing values of momentum transfer, as requested by the
 calculation of physical observables. In the Landau
 limit the q-range is restricted to $0\le q\le 2k^{\tau}_F(\rho)$, being $k^{\tau}_F$
 the Fermi momentum corresponding to the density $\rho_{\tau}$. In
 the present paper only $q = 0$ and $q = 2k_F$ will be discussed,
  $k_F$ being the Fermi momentum corresponding to the total density,.

The results are shown in Fig. 2, for the two values of the total spin
$S$ in ph coupling and the two limiting values of momentum transfer
range, $q=0$ and $q=2k_F$. Three values of the symmetry parameter
$\beta=(N-Z)/A$ have been chosen, including for comparison the
symmetric case. At $\beta=0$ the curves corresponding to n-n and p-p
matrix elements overlap, whereas at $\beta>0$ they split into two
different components for the isospin symmetry breaking in asymmetric
nuclear matter. The general trend is that ${\cal{J}}_i^S$ is much
larger at low density, as expected from the long-range character of
the induced interaction. In addition, the isospin splitting
${(\cal{J}}_i)_{pp}^S-{(\cal{J}}_i)_{nn}^S$ is increasing with
$\beta$, whereas ${(\cal{J}}_i)_{np}^S$ is almost insensitive. In
r.h.s. of Fig. 2 the dependence on the momentum transfer is depicted.
In our approximation only the kinematic dependence of the free
polarization propagator is considered, that is monotonically
decreasing with $q$. Therefore, as shown in Fig. 2, ${\cal{J}}_i$
takes the smallest values at the upper limit $q=2k_F$ of the
momentum transfer range.

In symmetric nuclear matter, $\beta=0$, the effective interaction can
be expressed in terms of the Landau-Migdal (LM) parameters
\cite{mig}. On microscopic basis the induced interaction
${(\cal{J}}_i)$ is the correction to the mean field prediction: the
S=0 curves correct the LM parameters $F_{nn}=F_{pp}$ and
$F_{np}=F_{pn}$, the S=1 curves correct the LM parameters
$G_{nn}=G_{pp}$ and $G_{np}=G_{pn}$. It is worthwhile noticing that,
in the induced interaction theory, the low-density instability
region of the equation of state, corresponding to negative
compression modulus, disappears at any $\beta$, as it can be easily
checked in the calculation of the LM parameters. This is  an old
standing result at $\beta=0$ \cite{back}.

The nuclear induced interaction has also been calculated in the spin
states of neutron-rich matter $(N>Z)$ in pp coupling, for the sake
of application to the NS pairing. The n-n and p-p matrix elements in
the spin singlet state are given by
\begin{equation}
{(\cal{J}}_i^{^1S_0})_{\tau,\tau} =\frac{1}{2}[
({{\cal{J}}_i)}^0_{\tau,\tau}\,-3\,({{\cal{J}}_i)}^1_{\tau,\tau}]\,\,,
\end{equation}
and in the spin-triplet pp state
\begin{equation}
{(\cal{J}}_i^{^3PF_1})_{\tau,\tau} = \frac{1}{2}[
({{\cal{J}}_i)}^0_{\tau,\tau}\,+\,({{\cal{J}}_i)}^1_{\tau,\tau}]\,\,.
\end{equation}
Moving from SNM to PNM the spin singlet interaction is mainly driven
by the interplay between density fluctuations ($S=0$) and spin
density fluctuations ($S=1$). The multiplicity of the latter  is
expected to play the major role considering that the two
contributions are of the same order of magnitude\cite{heisel}. The
spin triplet interaction is only driven by the interplay between
neutron and proton core polarizations.

\begin{figure}[htb]
\centering
\includegraphics[angle=0,width=15cm]{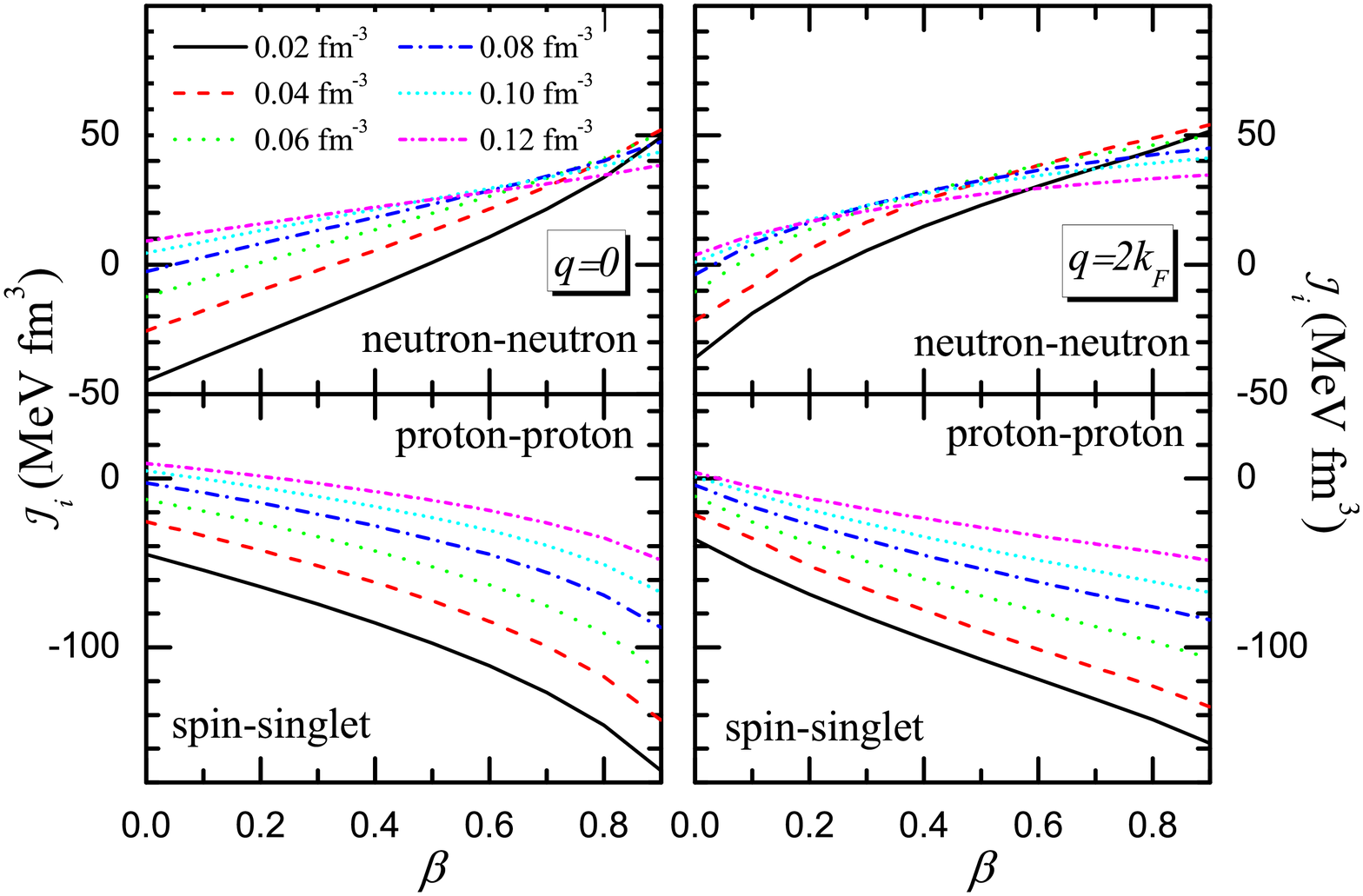}
\caption{Particle-particle induced interaction in the spin singlet
state at $q=0$ (left) and $q=2k_F$ (right). The upper (lower) panels
refer to neutron (proton) interaction.}
\end{figure}

 The results for the spin singlet state are displayed in Fig. 3 as a function
of the symmetry parameter for different densities of neutron-rich
matter. We notice that  the n-n effective interaction is attractive
on the side of symmetric nuclear matter, ($\beta=0$), indicating the
dominance of spin-density excitations over the density excitations,
whereas it is the other way around on the side of almost pure
neutron matter ($\beta=0.9$). Interesting is  the value of the
symmetry parameter $\beta_c$, where the two effects balance each
other, marking the crossover from attractive to repulsive
interaction. For asymmetries close to $\beta_c$ the effective
interaction is vanishing small, and, at variance with the symmetric
nuclear matter, no compensation is allowed between  self-energy and
 vertex corrections \cite{cao}. The value of the crossover
asymmetry $\beta_c$ is depending on density. There is density
threshold $\rho_c$: for $\rho > \rho_c$  the induced interaction is
repulsive at any asymmetry. In the present estimate for $\rho_c$ is
about 0.09 fm$^{-3}$. In the case of the p-p effective interaction
the situation is quite different, because in this case the neutron
core polarization is attractive and the proton core polarization is
repulsive. Therefore, increasing asymmetry, the former  becomes more
and more dominant, resulting into  an  increasingly attractive
strength of the induced interaction. In the p-p case the critical
density $\rho_c$ is the lower limit for the existence of the isospin
crossover.

 The results at $q = 2k_F$ are depicted in r.h.s. of Fig. 3.
 From the kinematic behavior of the free propagator it turns out that, increasing
 the momentum transfer, the crossover is pushed to lower values of both asymmetry and density
 in the n-n case but rapidly disappears in the case of protons.

 The p-p effective interaction in spin singlet state was studied in
 Ref. \cite{baldo} for nuclear matter in $\beta$-stable regime. At $\rho$=0.2 fm$^{-3}$ it was
 found a value of -14.2 MeV fm$^{3}$  to be compared with
 the value -1.2 MeV fm$^{3}$ from the present study. The deviation by one order of magnitude
 should not be considered too large, taking into account the different
 approximations adopted in the two approaches, in particular, the missing three-body force
 in the Ref. \cite{baldo} calculation.

 The results for the triplet case are depicted in Fig. 4. In such a case
 there no interplay between density and spin density fluctuations so
 that no any crossover is possible. The isospin dependence is easily
 understood by the competition  between neutron
 and proton polarization: moving from SNM to PNM the n-n ${\cal
 J}_i$ monotonically decreasing and p-p ${\cal J}_i$ is
 monotonically increasing. The main feature is that the induced interaction  is always
 repulsive, the repulsion being attenuated by finite momentum transfer, as shown in Fig. 4 (left).
 This property may eventually explain why in finite nuclei it is found
that strength of T=0 pairing is never stronger than T=1 strength in
spite of the fact that the bare T=0 force yields a much stronger gap
than the S=0 one.
\begin{figure}[htb]
\centering

\includegraphics[angle=0,width=15cm]{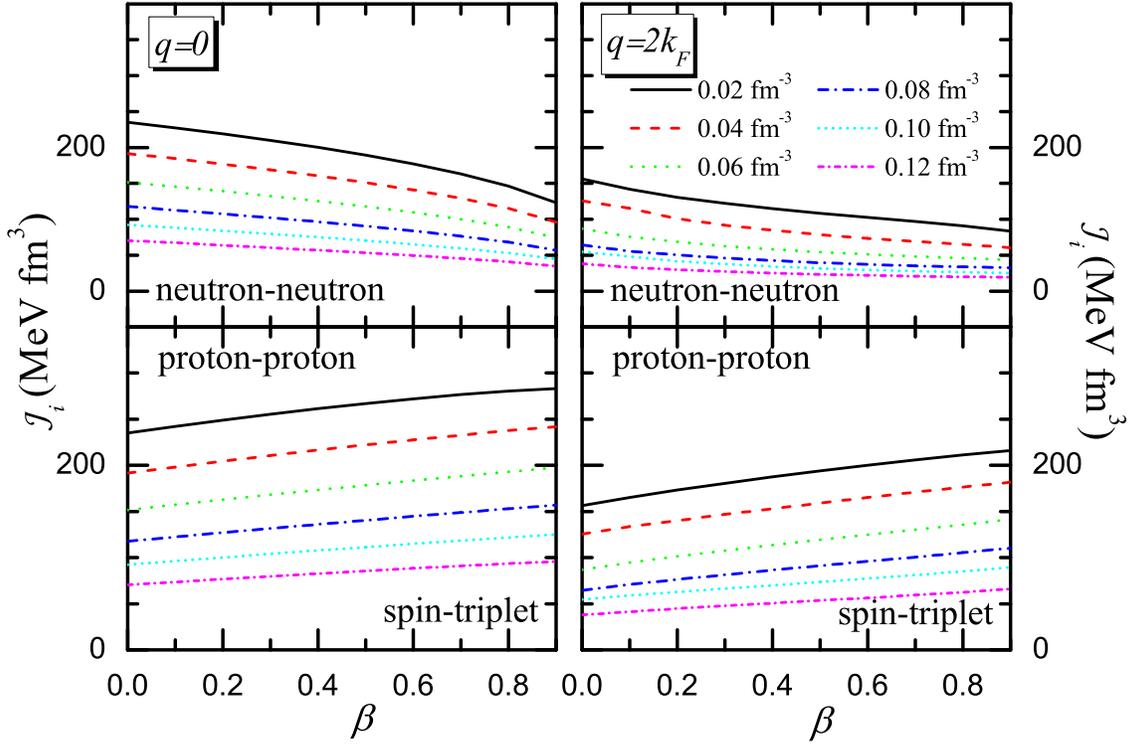}

\caption{Particle-particle induced interaction in the spin triplet
state at $q=0$ (left) and $q = 2k_F$ (right). The upper (lower)
panels refer to neutron (proton)
interaction.}
\end{figure}

In conclusion, from the calculations of the induced interaction one
should expect the core polarization to  have a deep influence on the
pairing interaction in $\beta$-stable nuclear matter and, in
particular, on the three superfluid states supposed to exist in NS.
In the case of the low-density n-n  $^1S_0$ pairing, located in the
NS crust, the induced interaction turns out to be repulsive, as
shown in the Fig. 3 for spin singlet states, because the crust is a
very neutron-rich state. Therefore, the gap is expected to be
reduced not only by the self-energy corrections, but also by the
induced interaction. On the other hand, in the case of the high
density $^3PF_2$ n-n pairing, located in the NS inner core, the
strong core polarization, that is always repulsive in spin triplet
states, should suppress completely the high density n-n pairing, at
variance with Ref. \cite{dickhoff}. On the contrary, in the case of
$^1S_0$ p-p pairing, where low-density proton fraction is embedded
into high-density neutron matter, the neutron core induces a strong
attractive enhancement on the p-p Cooper pairs so to compete with
the self-energy suppression.

Therefore, after the introduction  in the gap equation of the
self-energy corrections of $\beta$-stable nuclear matter
\cite{dong}, the inclusion of the induced interaction on the same
footing should rise the theoretical study of NS superfluidity to a
quite satisfactory stage. Such calculations are in progress.

\begin{acknowledgments}This paper is partially  supported  by the National
Natural Science Foundation of China under Grant Nos 11375022, 11175216, 11575060,
and 11435014, 11235002 and the Fundamental Research Funds for the Central Universities
(JB2014241).
\end{acknowledgments}

\end{document}